\documentclass[12pt]{iopart} 

\usepackage{amssymb}
\usepackage{epsfig}

\bibstyle{unsrt}

\newcommand{\bra}[1]{\langle{#1}|}
\newcommand{\ket}[1]{|{#1}\rangle}
\def\ketc[#1]{\vert #1 \rangle}
\def\brac[#1]{\langle #1 \vert}

\newcommand{\beq}{\begin{equation}}
\newcommand{\eeq}{\end{equation}}
\newcommand{\bqa}{\begin{eqnarray}}
\newcommand{\eqa}{\end{eqnarray}}
\newcommand{\nn}{\nonumber}

\newcommand{\erf}[1]{Eq.~(\ref{#1})}
\newcommand{\dg}{^\dagger}

\begin{document}

\title{Optimal estimation of one parameter quantum channels}

\author{Mohan Sarovar and G J Milburn}
\address{Centre for Quantum Computer Technology, and School of Physical Sciences, The
University of Queensland, St Lucia, QLD 4072, Australia}
\eads{\mailto{mohan@physics.uq.edu.au}, \mailto{milburn@physics.uq.edu.au}}


\begin{abstract}
We explore the task of optimal quantum channel identification, and in particular the estimation of a general one parameter quantum process. We derive new characterizations of optimality and apply the results to several examples including the qubit depolarizing channel and the harmonic oscillator damping channel. We also discuss the geometry of the problem and illustrate the usefulness of using entanglement in process estimation.
\end{abstract}

\pacs{03.65.Ta, 03.67.-a}
\submitto{\JPA}

\newtheorem{attainability}{Theorem}
\newtheorem{uniqueness}[attainability]{Theorem}

\maketitle


\section{Introduction}
\label{sec:intro}

\begin{figure}[h!]
\centering
\leavevmode
\includegraphics[scale=0.6]{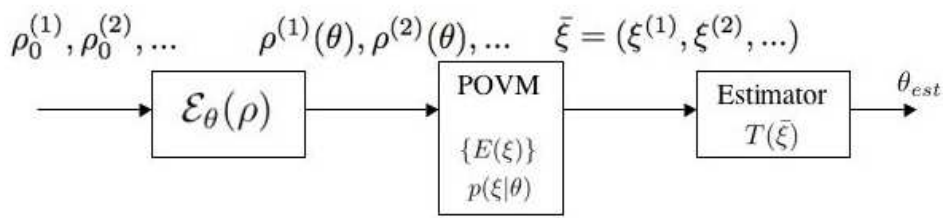}
\caption{The estimation procedure} \label{fig:schematic}
\end{figure}

Recent experimental progress in quantum information processing has highlighted the importance of quantum control and in particular, the task of system identification. For example, it is essential for the verification of quantum gates to be able to effectively identify quantum processes. This identification might require a full process tomography, but quite often an estimation of a number of free parameters may be sufficient to determine the proper operation of a gate. In such cases, it is obviously beneficial to use optimal schemes to estimate the free parameters. With that motivation we consider the problem of optimally estimating a quantum process with one free parameter. Figure \ref{fig:schematic} represents a schematic of the problem. The goal is to identify an input state and a measurement scheme that will permit one to gain the most informaiton about the free parameter. We will make these notions more precise in what follows.

The field of quantum parameter estimation is concerned with the methods of estimating -- especially optimally estimating -- properties of quantum states or processes. The field has a fairly short but rich history, beginning with the pioneering work of Helstrom \cite{Hel-1976} and Holevo \cite{Hol-1982} and has recently seen a rise in interest, especially from the quantum information community. 

The task of estimating quantum states has an enormous literature dedicated to it (a small sample is \cite{Bra.Cav-1994, Fuj.Nag-1995, Bar.Gil-2000, Gil.Mas-2000, Hol-2004}, and for a recent review see Ref. \cite{Par.Reh-2004}). There are several notions of optimality in this scenario and some of them highlight the essential connection between the optimal estimation problem and the geometry of the space of quantum states \cite{Woo-1981, Bra.Cav-1994}: the statistical distingushability of states induces a unique metric (the \textit{Fisher metric}) on the space of quantum states \cite{Bra.Cav-1994, Bar.Gil.etal-2003}. This is a pleasing state of affairs, which although not of much practical use (it cannot be used to identify optimal estimation strategies except in some special cases \cite{Bra.Cav.etal-1994}), is tremendously edifying. 

In contrast, parameter estimation for quantum processes is an area that is far less developed. The estimation of unitary quantum processes has been examined by a number of authors over the past few decades \cite{Hel-1976, Hol-1982, Bra.Cav.etal-1994, Fuj.Nag-1995, Bal-2004}; in these treatments, the system evolves unitarily and the parameters to be estimated are unknown parameters of this unitary. However, the general form of the problem, where the evolution is a general quantum operation, has received considerably less attention. Some recent treatments of special cases of this more general problem are Refs. \cite{Fuj-2001, Fuj.Hir-2003, Sas.Ban.etal-2002}. 

One way to attack the problem of optimal quantum process estimation is to treat it as an optimal state estimation problem where the states under consideration are restricted to the parametrized (by the parameters of the process) set of output states of the quantum process. This is an practical point of view which recognizes that the only operational access to the parameters of the process is through probe input states that are measured at the output of the process (see Fig. \ref{fig:schematic}). This approach breaks the process parameter estimation into two parts: the choice of an optimal input state, and the choice of an optimal estimation scheme for the parametric family of output states $\{\mathcal{E}_\theta (\rho_0)\}_\theta$. Such an approach has been taken in the past (e.g. \cite{Fuj-2001}) but with the dynamics of the process (or equivalently, the dependency of the output parameteric family on the parameters of the process) being represented by rather abstract superoperators such as the \textit{symmetric logarithmic derivative} \cite{Bra.Cav-1994}. The disadvantage of using such a representation is that explicit expressions for such superoperators are often difficult to calculate. 

The aim of this paper is to use the same approach to the estimation of quantum processes, but to use a more common representation for them. In particular, we assume that the Kraus representation (operator sum decomposition) of the quantum process is given, with the parameters to be estimated being free parameters of the Kraus operators. We consider the simplest quantum process estimation problem in this general setting: the estimation of a one parameter, trace-preserving quantum operation; and investigate the advantages and disadvantages of using the Kraus representation to describe the process. We assume that the input state remains fixed and optimize over the estimation scheme to arrive at a measure of optimality. The measure is non-unique and non-constructive (it does not allow one to deduce the optimal POVM), primarily due to the non-uniqueness of the Kraus representation. However, for a special family of quantum channels we show that it can be used to test the optimality of an estimation scheme. We also discuss the geometry of the problem and apply the results to several examples, including one which illustrates the value of using entanglement to increase the statistical distinguishability of quantum processes. 

The paper is organized as follows: section \ref{sec:prelim} covers some preliminary concepts and sets the notation. Section \ref{sec:opt} derives the optimality conditions for the estimation scheme and discusses them. Then we consider several examples in section \ref{sec:examples}, and finally conclude in section \ref{sec:concl}.

\section{Preliminaries}
\label{sec:prelim}

\subsection{Quantum operations}
\label{sec:qops}
Given two Hilbert spaces, $\mathcal{H}_1$ and $\mathcal{H}_2$, with dimensions $d_1$ and $d_2$ respectively, a quantum operation, $\mathcal{E}: \mathcal{H}^T_1 \rightarrow \mathcal{H}^T_2$ is a \textit{completely positive} linear map between trace-class operators in $\mathcal{H}_1$ and $\mathcal{H}_2$ ($\mathcal{H}^T_i$ is the space of trace-class operators acting on Hilbert space $\mathcal{H}_i$; $\textrm{dim}\mathcal{H}^T_i = d_i^2$). These maps represent the most general transformations between two density operators in quantum mechanics. The complete positivity of the map is a physically motivated and highly restrictive condition, however, there is a well known representation of such maps originally due to Choi \cite{Cho-1975} and then popularized by Kraus \cite{Kra-1983}, called the \textit{operator sum decomposition}, or \textit{Kraus decomposition}. This decomposition represents the action of the map $\mathcal{E}: \mathcal{H}^T_1 \rightarrow \mathcal{H}^T_2$ as 
\begin{equation}
\label{eq:kraus_decomp}
\tilde{\rho} \equiv \mathcal{E}(\rho) = \sum_k \Upsilon_k \rho \Upsilon\dg_k
\end{equation}
where $\rho \in \mathcal{H}^T_1$ and $\tilde{\rho} \in \mathcal{H}^T_2$. The operators $\Upsilon_k: \mathcal{H}_1 \rightarrow \mathcal{H}_2$ are called the \textit{Kraus operators} of the decomposition. The sum in this decomposition can run over an infinite set, but in general $\mathcal{E}$ can be represented using at most $d_1d_2$ Kraus operators \cite{mikeandike}. In this paper we will restrict our attention to \textit{trace-preserving} quantum operations between isomorphic Hilbert spaces (i.e. $d_1 = d_2 = d$). The trace preservation implies that the Kraus operators satisfy a normalization condition:
\begin{equation}
\sum_k \Upsilon_k\dg \Upsilon_k = I_d
\end{equation}
where $I_d$ is the d-dimensional identity operator. Of course, for a closed quantum system undergoing unitary evolution, there is only one Kraus operator in the decomposition and it is unitary - i.e. $\mathcal{E}(\rho) = V\rho V\dg$ with $V\dg V = I_d$.

An important property of this decomposition is the non-uniqueness of the Kraus operators. A new set of Kraus operators for the same quantum operation can be derived by an arbitrary unitary remixing of the original set. That is, if $\{\Upsilon_k\}_1^N$ are the Kraus operators in a decomposition of $\mathcal{E}$, then the set $\{\Omega_j\}_1^N$:
\begin{equation}
\Omega_j = \sum_{k=1}^N u_{jk}\Upsilon_k
\end{equation}
where $u_{jk}$ are the elements of a unitary matrix, are the Kraus operators for a different, but equivalent, operator sum decomposition of $\mathcal{E}$. This non-uniqueness will become important when we describe the estimation of quantum operations in terms of their Kraus operators; in particular, a specific decomposition, called the \textit{canonical decomposition}, will be important. The canonical decomposition is a distinguished Kraus decomposition that can be constructed \textit{with respect to any given input state} with the following property: $\mathcal{E}(\rho) = \sum_k \Upsilon_k\rho \Upsilon_k\dg$ with $\tr( \Upsilon_k\dg \Upsilon_j \rho) = \delta_{jk} p_k$. Trace preservation implies $\sum_k p_k = 1$. The explicit construction of these Kraus operators is detailed in Ref. \cite{Nie.Cav.etal-1998}, however for our purposes the important thing to notice is their dependence on the input state -- if the input state changes, the Kraus operators in the canonical decomposition for a given quantum process will change.

Now we describe an important interpretation of quantum operations. By the postulates of quantum mechanics, a closed quantum system will undergo unitary evolution. Any departure from such an evolution is caused by coupling of the system to additional degrees of freedom (usually termed the environment). Therefore, any non-unitary quantum operation is the effective description of the evolution of a quantum system that is coupled to an environment. This leads us to think of a quantum operation, acting on a system of interest, as a unitary map of the system plus some environment (which combined form a closed system) after which the environment is traced out \cite{mikeandike}. That is,
\begin{equation}
\mathcal{E}(\rho) =
\tr_{env}[U(\rho\otimes\rho_{env})U\dg]
\end{equation}
where $\rho$ is a density operator for the system, and $\rho_{env}$ is a density operator for the environment. $U$ is a unitary operator acting on both the system and the environment. This is sometimes referred to as the \textit{dilation} of the quantum operation. Now, assuming that $\{\ket{e_k}\}$ is a complete orthonormal basis for the state space of the environment and
the initial environment state is $\rho_{env} = \ket{e_0}\bra{e_0}$, then the Kraus decomposition of $\mathcal{E}$ can be written explicitly by performing the trace in this basis:
\begin{eqnarray}
\mathcal{E}(\rho) &=&
\tr_{env}[U(\rho \otimes \rho_{env})U\dg] \nn \\
 &=& \sum_k \bra{e_k}U ~[\rho \otimes
\ket{e_0}\bra{e_0}]~ U\dg \ket{e_k} \nn \\
 &=& \sum_k \Upsilon_k \rho \Upsilon\dg_k
\end{eqnarray}
where $\Upsilon_k = \bra{e_k}U\ket{e_0}$ is an operator acting only on the system subspace. The unitary freedom in the choice of Kraus operators is exactly the same as the unitary freedom in choosing the environmental basis states in which to perform the trace. The assumption of a pure initial state for the environment is not very restrictive because we can always choose the environment to be large enough so that this condition is met. The other assumption, that the initial state is separable, is a much more subtle one and a full discussion of it is out of the scope of this paper. It has been extensively discussed in the literature, and we refer the interested reader to the recent treatment in Ref. \cite{Ben.Flo-2005}.

\subsection{Optimal estimation}
\label{sec:opt_est}
We are concerned with optimally estimating a one-parameter quantum operation, $\mathcal{E}_\theta$. We assume that we have available an operator sum decomposition of the operation, but that the Kraus operators have a free \textit{real, continuous} parameter, $\theta$, which is to be estimated. 
\begin{eqnarray}
\rho(\theta) \equiv \mathcal{E}_\theta(\rho_0) = \sum_{k}
\Upsilon_k(\theta) \rho_0 \Upsilon_k\dg(\theta)
\end{eqnarray}
with $\sum_{k} \Upsilon_k(\theta)\dg \Upsilon_k(\theta) = I_d$. Here $\rho_0 \in \mathcal{H}^T ~ (\dim\mathcal{H}^T=d^2)$ is the input state to the operation over which we have complete control. A schematic of the estimation process is given in Figure \ref{fig:schematic}. In each run of the experiment, the input state $\rho_0$ is fed into the quantum operation and a generalized measurement is performed on the output. The generalized measurement is described by non-negative, Hermitian operators $E(\xi)$, or POVMs, which satisfy the completeness property: $\int d\xi E(\xi) = I_d$. $\xi$ labels the result/s of the measurement and can be univariate or multivariate, as well as continuous or discrete (in the discrete case, the completeness integral becomes a sum). Given such a description of the measurement, the probability density for the measurement result, conditioned on a given state is given by $p(\xi|\theta) = \textrm{tr}(E(\xi)\rho(\theta))$. We will assume that there are $N$ independent runs of this experiment after which the parameter $\theta$ is estimated using the results of $N$ independent, seperable measurements on $N$ outputs from the channel. That is, $\theta_{est} = \theta_{est}(\xi_1, \xi_2, ..., \xi_N)$. The estimator, $T$ is any function of the $N$ measurement outcomes, and it attempts to reconstruct $\theta$ from these outcomes. We will assume that the estimator, $T$, is unbiased. That is, $E_\theta \{\theta_{est}\} = \theta$ where $E_\theta \{.\}$ is the expectation value with respect to the probability distribution for $\theta$. This is a mild assumption that will not affect the essential results. An example of an estimator (which is unbiased also) is the \textit{sample mean}: $\bar{m} = 1/N \sum_{i=1}^{N} x_i$, which is an estimate of the true mean of the probability distribution that the $x_i$ are drawn from.

This is a very general setting in which to describe parameter estimation. We can account for scenarios such as entanglement assisted estimation by simply changing the definition of the quantum operation to be a tensor product of two or more operations. We will see an example of this in section \ref{sec:examples}. We are seeking the optimal scheme for estimating the free parameter, $\theta$. Loosely, this amounts to specifying an input state, a measurement scheme, and an estimator that will permit one to gain the most information about $\theta$, and we will now proceed by making this notion of optimality more precise. 

Firstly, the deviation of our estimate from the actual value of the parameter can be measured by:
\begin{equation}
\label{eq:stat_error}
\delta\theta \equiv \theta_{est} - \theta
\end{equation}
It is natural to consider the estimation scheme that minimizes the variance of this estimation error as the optimal one. That is, we want to minimize $\langle (\delta\theta)^2 \rangle$. Braunstein and Caves consider a similar problem in Ref. \cite{Bra.Cav-1994}, and we will follow their treatment in order to find the optimal scheme. 

As mentioned in the introduction, this optimization problem can be split into two subproblems: (i) the choice of an optimal input state $\rho_0$, and, (ii) the choice of an optimal scheme (choice of $\{E(\xi)\}$ and $T$) to estimate from the one parameter family of output states $\rho (\theta)=\mathcal{E}(\rho_0)$. In this paper, we will not consider the first subproblem -- we will assume that the input state is fixed and focus on the optimization of the estimation scheme. The second subproblem, the optimization of the estimation scheme can be further broken down into two steps: first, a minimization of the error variance over estimators $T$ for a given quantum measurement, and second, a minimization over all quantum measurements. The optimization over estimators is an entirely classical one - it is well known in the statistical inference literature and results in the famous Cram\'{e}r-Rao bound \cite{Cov.Tho-1991}:
\begin{equation}
\label{eq:cramer_rao}
\langle (\delta\theta)^2 \rangle \geq \frac{1}{NF(\theta)}
\end{equation}
where $N$ is the number of measurement results used in the estimation, and $F(\theta)$ is the \textit{Fisher information}:
\begin{eqnarray}
\label{eq:fisher_info}
F(\theta) &\equiv& \int d\xi p(\xi|\theta) \left( \frac{ \partial \ln p(\xi |\theta)}{ \partial \theta} \right) ^2 \nn \\
&=&  \int d\xi \frac{1}{p(\xi|\theta)} \left( \frac{ \partial p(\xi |\theta)} { \partial \theta} \right) ^2
\end{eqnarray}
The Fisher information represents the amount of information about $\theta$ contained in the measurement result $\xi$. The dependence of this quantity on the choice of quantum measurement is clear from the fact that $p(\xi | \theta) = \textrm{tr}(E(\xi)\rho(\theta))$. Strictly, this form of the bound is only valid for unbiased estimators. As we will only consider such estimators we will not state the more general form of the bound here. 

The Cram\'{e}r-Rao bound effectively takes the estimator out of the picture. It says that for a given input state and measurement (i.e. for a given $p(\xi |\theta)$) the variance is lower bounded by the the quantity on the right-hand side of \erf{eq:cramer_rao}. Thus subproblem (ii) -- the estimation on the parametric family of output states -- simplifies to finding the best measurement: the one that minimizes this lower bound, or equivalently, maximizes the Fisher information. As an aside, under mild regularity conditions on $p(\xi |\theta)$ the lower bound in \erf{eq:cramer_rao} is an asymptotically achievable one; that is, there exist estimators that can attain this bound as $N \rightarrow \infty$, and an example is the maximum likelihood estimator \cite{Nah-1969}. The estimators that achieve this bound have been extensively studied in the field of statistical inference, therefore we will not consider them here but will rather concentrate on the quantum aspects of the problem.

Given the Cram\'{e}r-Rao bound, the next step of the optimization becomes a maximization of \erf{eq:fisher_info} over all possible quantum measurements. We notate this maximization, and the result by:
\begin{equation}
\label{eq:task}
F^{*}(\theta) \equiv \max_{\{E(\xi)\}} F(\theta)
\end{equation}

\subsubsection{A geometric perspective}
\label{sec:geom}
Before treating this maximization, we examine the process estimation problem from a geometric perspective. The one parameter family formed by the output of the process for a fixed input state, $\{\rho(\theta)\}_\theta$, defines a curve in density operator space which is parametrized by the continuous, real parameter $\theta$. The curve is itself a manifold defined by $\rho_0 \equiv \ket{\psi_0}\bra{\psi_0}$ and $\mathcal{E}_\theta$, and the advantage of regarding members of $\{\rho(\theta)\}_\theta$ as outputs of a quantum process represented by its Kraus decomposition is that we can define a natural local co-ordinate patch at each point on the curve \footnote{Here we have assumed that the input state is pure. We will see below that the optimal input state will always be a pure one, and therefore this assumption is justified.}. That is, $\rho(\theta) = \sum_k \Upsilon_k(\theta) \ket{\psi_0}\bra{\psi_0} \Upsilon_k\dg (\theta) = \sum_k \ket{e_k(\theta)}\bra{e_k(\theta)}$, where $\ket{e_k(\theta)} \equiv \Upsilon_k(\theta)\ket{\psi_0}$ are unnormalized vectors. Now if we exclusively use the canonical decomposition for the process, this can be rewritten as an eigendecomposition of $\rho(\theta)$: $\sum_k p_k(\theta) \ket{f_k(\theta)}\bra{f_k(\theta)}$, where $\ket{f_k(\theta)} = \frac{1}{\sqrt{p_k(\theta)}}\ket{e_k(\theta)}$ and $\langle f_j(\theta) \vert f_k(\theta) \rangle = \delta_{jk}$. The set $\{\ket{f_k (\theta)}\}$ can be considered a local orthonormal co-ordinate basis at the point $\theta$ on the curve. A point to note is that while we write the eigendecomposition as a sum over $k$, the number of Kraus operators in the canonical decomposition, it does not mean that $\rho(\theta)$ has the same number of eigenvectors as the number of Kraus operators there are in the Kraus decomposition. $\rho(\theta)$ can have at most $d$ eigenvectors, where $d$ is the dimension of the system, while there is no limit on the number of Kraus operators. In the canonical decomposition, $\Upsilon_k(\theta)\ket{\psi_0} = 0$ for some values of $k$, and these terms will drop out in the eigendecomposition sum.

The Fisher information can be used to define a Riemannian metric on this curve (submanifold), that measures the statistical distinguishability of neighbouring one-parameter quantum operations given the fixed input state $\ket{\psi_0}$. To see this, we go back to the definitions above and note that the Cram\'{e}r-Rao bound \erf{eq:cramer_rao} is a lower bound on the variance in the error when estimating the parameter $\theta$. Thus it is a lower bound on the error in reliably distinguishing between two neighbouring quantum operations: $\mathcal{E}_\theta$ and $\mathcal{E}_{\theta+d\theta}$. Therefore, as in \cite{Bra.Cav-1994}, we can consider it a \textit{distinguishability metric} along the curve of one parameter quantum operations defined by $\ket{\psi_0}$ and $\mathcal{E}_\theta$. More formally, let us establish $\textrm{min}[\sqrt{N}\langle (\delta\theta)^2 \rangle^{1/2}]$ to be a measure of statistical deviation. The $\sqrt{N}$ removes the improvement in estimation due to increased sampling, and the minimization is over measurement schemes to ensure that we are considering the most discriminating scheme. A statistical measure of distinguishability should be proportional to the inverse of this deviation measure - i.e. the more the deviation, the less distinguishable neighbouring operations become. Thus we can define a distinguishability metric along the curve as:
\begin{equation}
ds^2 = \frac{d\theta^2}{\textrm{min}[N\langle (\delta\theta)^2 \rangle]}
\end{equation}
$(ds/d\theta)^2$ is well known as the \textit{statistical distance} \cite{Bra.Cav-1994}, and using \erf{eq:cramer_rao}, we can rewrite it in terms of the Fisher information as
\begin{equation}
\label{eq:stat_dist}
\left( \frac{ds}{d\theta} \right)^2 = \max_{\{E(\xi)\}}  F(\theta) = F^*(\theta)
\end{equation}
This is exactly the maximization we are considering for optimal estimation. Note that it is only over the measurement POVMs ($\{E(\xi)\}$) because this is the statistical distance over a curve defined by a particular input state. A caveat is required here: when we refer to $F^*(\theta)$ as being the metric on the curve, this is strictly only true if it can be shown that the bound set by $F^*(\theta)$ can be achieved. This question of achievability will be important in the following. 

\section{The optimization}
\label{sec:opt}

As outlined in section \ref{sec:opt_est}, the procedure of finding the best estimation scheme can be phrased as a sequence of optimizations. The first optimization, which is entirely classical, results in the Cram\'{e}r-Rao bound, and in this section we shall examine the quantum aspects of the problem. 

\subsection{Optimal estimation on the output family}
\label{sec:opt_meas}
The optimal quantum measurement scheme is the set of POVMs that maximize the Fisher information, for a fixed input state. Using the definition of the Fisher information \erf{eq:fisher_info}, and the fact that $p(\xi | \theta) = \textrm{tr}(E(\xi)\rho(\theta))$ we get:  
\begin{eqnarray}
\label{eq:q_fish}
F^*(\theta) = \max_{\{E(\xi)\}} \int d\xi \frac{(\tr[E(\xi)\rho
~'(\theta)])^2}{\tr[E(\xi)\rho(\theta)]}
\end{eqnarray}
where $\rho(\theta) \equiv \mathcal{E}_\theta(\rho_0) = \sum_{k}
\Upsilon_k(\theta) \rho_0 \Upsilon_k\dg(\theta)$ and $\rho ~'(\theta) = \partial \rho(\theta)/\partial \theta$. Let $\{\Upsilon_k\}$ be the Kraus operators for an arbitrary Kraus decomposition of $\mathcal{E}$. Now, the next logical step would be to replace $\rho(\theta)$ and $\rho~'(\theta)$ by their definitions in terms of the Kraus operators that define the quantum operation. However, this makes the maximization of (\ref{eq:q_fish}) difficult due to the introduction of the Kraus decomposition sum in the numerator. Instead, we will take a step back and use the dilation of the quantum operation. 

As mentioned in section \ref{sec:qops}, a quantum operation can be thought of as a unitary map of the system plus some environment after which the environment is traced out. Given this, we will label our system A and the environment B, and define:
\begin{eqnarray}
\label{eq:rho}
 \rho_A(\theta) &=& \mathcal{E}_\theta (\rho_A^0)
= \sum_k \Upsilon_k(\theta) \rho_A^0 \Upsilon\dg_k(\theta) \nn \\
&=& \tr_B \{~ U(\theta) ~[\rho_A^0 \otimes
\ket{e_0}_B\bra{e_0}]~ U\dg(\theta) ~\}
\end{eqnarray}
where $U(\theta)$ is some unitary operator acting on systems A and
B, and $\rho_A^0$ is the input state on subsystem A. The mapping between $U(\theta)$ and $\{\Upsilon_k(\theta)\}$ is not unique because of the freedom in choosing the environment basis states, and we will return to this point shortly. Also,
\begin{eqnarray}
\label{eq:d_rho} \rho_A '(\theta) &=& \tr_B \{~  U~'(\theta)
~[\rho_A^0 \otimes \ket{e_0}_B\bra{e_0}]~ U\dg(\theta) \nn \\
&+& U(\theta) ~[\rho_A^0 \otimes \ket{e_0}_B\bra{e_0}]~
{U\dg}~' (\theta) ~\} \nn \\
&=& \tr_B \{~ \Omega(\theta) + \Omega\dg(\theta) ~\}
\end{eqnarray}
where $U'(\theta) = \partial U(\theta)/\partial \theta$, and
$\Omega(\theta) = U~'(\theta) ~[\rho_A^0 \otimes
\ket{e_0}_B\bra{e_0}]~ U\dg(\theta)$.

Now we return to the problem of maximizing (\ref{eq:q_fish}). Substituting (\ref{eq:rho}) and (\ref{eq:d_rho}), we get (in the following, we will use $E_A(\xi) \otimes I_B$ and $E_A(\xi)I_B$ interchangeably to denote the same operator):
\begin{eqnarray}
&&\int d\xi ~\frac{ (\tr_A \{~ E_A(\xi) \tr_B
\{~\Omega(\theta) + \Omega\dg(\theta)~\}~\})^2}{\tr_A
\{~E_A(\xi) \tr_B \{~ U(\theta) ~[\rho_A^0 \otimes
\ket{e_0}_B\bra{e_0}]~ U\dg(\theta) ~\}\}} \nn \\
&=& \int d\xi ~\frac{ (\tr_A \{~ \tr_B \{~ (E_A(\xi) I_B)(\Omega(\theta) + \Omega\dg(\theta))~\}~\})^2}{\tr_A \{~\tr_B \{~ (E_A(\xi) I_B) U(\theta) ~[\rho_A^0 \otimes \ket{e_0}_B\bra{e_0}]~ U\dg(\theta) ~\}\}} \nn \\
&=& \int d\xi ~\frac{ (\tr \{~ (E_A(\xi) I_B)\Omega(\theta) + (E_A(\xi) I_B)\Omega\dg(\theta)~\})^2}{\tr \{~ (E_A(\xi) I_B) U(\theta) ~[\rho_A^0 \otimes \ket{e_0}_B\bra{e_0}]~ U\dg(\theta) ~\}} \nn \\
&=& 4\int d\xi ~\frac{ ( \Re ~ \tr \{~ (E_A(\xi) I_B)\Omega(\theta)~\} )^2}{\tr \{~ (E_A(\xi) I_B)
U(\theta) ~[\rho_A^0 \otimes \ket{e_0}_B\bra{e_0}]~ U\dg(\theta) ~\}} \nn \\
&\leq& 4\int d\xi ~\frac{ | ~ \tr \{~ (E_A(\xi) I_B)\Omega(\theta)~\} |^2}{\tr \{~ (E_A(\xi) I_B)
U(\theta) ~[\rho_A^0 \otimes \ket{e_0}_B\bra{e_0}]~
U\dg(\theta) ~\}} \nn \\
&=& 4\int d\xi ~\frac{ | ~ \tr \{~ (E_A(\xi) \otimes I_B) U~'(\theta) ~[\rho_A^0 \otimes
\ket{e_0}_B\bra{e_0}]~ U\dg(\theta) ~\} |^2}{\tr \{~(E_A(\xi) \otimes I_B) U(\theta) ~[\rho_A^0 \otimes \ket{e_0}_B\bra{e_0}] ~U\dg(\theta) ~\}} \nn
\end{eqnarray}
We proceed by applying the Cauchy-Schwarz inequality: $|\tr(O\dg P)|^2 \leq \tr(O\dg O)\tr(P\dg P)$, with equality when $O=\lambda P$ for some constant $\lambda$. We will apply this inequality to the numerator, where $O\dg = (E_A(\xi)I_B)^{1/2} U~'(\theta) ~[\rho_A^0 \otimes \ket{e_0}_B\bra{e_0}]^{1/2}$ and $P = [\rho_A^0 \otimes \ket{e_0}_B\bra{e_0}]^{1/2}~ U\dg(\theta) (E_A(\xi)I_B)^{1/2}$, to get:

\begin{eqnarray}
F(\theta) &\leq& 4\int d\xi ~\frac{ \tr \{~ (E_A(\xi)I_B) U(\theta) ~[\rho_A^0 \otimes \ket{e_0}_B\bra{e_0}] U\dg(\theta) ~\}}{\tr \{~(E_A(\xi)I_B) U(\theta)~[\rho_A^0 \otimes \ket{e_0}_B\bra{e_0}]~ U\dg(\theta) ~\}} \nn \\
&& ~~~~~~~~ \times \tr \{~ (E_A(\xi)I_B)
U~'(\theta) ~[\rho_A^0 \otimes \ket{e_0}_B\bra{e_0}]~ {U\dg}'(\theta) ~\} \nn \\
&=& 4\int d\xi ~\tr \{~ (E_A(\xi) \otimes I_B) U~'(\theta)
[\rho_A^0 \otimes \ket{e_0}_B\bra{e_0}]~ {U\dg}'(\theta) ~\} \nn \\
&=& 4\tr \{~ U~'(\theta) ~[\rho_A^0 \otimes \ket{e_0}_B\bra{e_0}] ~{U\dg}'(\theta) ~\} \nn \\
&=& 4\tr_A\tr_B \{~ U~'(\theta) ~[\rho_A^0 \otimes \ket{e_0}_B\bra{e_0}]~ {U\dg}'(\theta) ~\} \nn \\
&=& 4\tr_A \{~ \sum_k \Upsilon_k '(\theta) \rho_A^0 {\Upsilon_k\dg}'(\theta) ~\} \nn \\
&=& 4 \tr \{ \sum_k {\Upsilon_k\dg}'(\theta) \Upsilon_k '(\theta) ~\rho_0 \}
\end{eqnarray}
where we have dropped all subscripts in the last line because all operators are in subsystem A, and the completeness relation $\int d\xi E(\xi) = I_d$ has been used.

We have arrived at a bound on the Fisher information:
\begin{equation}
\label{eq:c_up}
C_\Upsilon(\theta) \equiv 4 \tr \{ \sum_k {\Upsilon_k\dg}'(\theta) \Upsilon_k '(\theta) ~\rho_0 \} \end{equation}
This bound is equal to the maximum, $F^*(\theta)$, \textit{if} it is achievable. To achieve the bound we need to saturate the two inequalities used in the derivation.
The condition for the meeting the first inequality is:
\begin{eqnarray}
\label{eq:cond1}
\Im ~ \tr \{~ (E_A(\xi) \otimes I_B)\Omega(\theta)~\} = 0
\nn \\ 
\iff \Im ~ \tr_A \{~ E(\xi) \sum_k \Upsilon_k'(\theta)
~\rho_0~\Upsilon_k\dg(\theta)~\} = 0 \nn \\
\iff \Im ~ \tr \{\sum_k \Upsilon_k\dg(\theta) E(\xi) \Upsilon_k'(\theta) ~\rho_0\} = 0 ~~~ \forall \xi 
\end{eqnarray}

The condition for meeting the Cauchy-Schwarz bound is
\begin{eqnarray}
&& (E_A(\xi)\otimes I_B)^{\frac{1}{2}} U~'(\theta) ~[\rho_A^0 \otimes \ket{e_0}_B\bra{e_0}]^{\frac{1}{2}} \nn \\
&& = ~~~~~ \lambda_\xi (\theta) ~(E_A(\xi)\otimes I_B)^{\frac{1}{2}} U(\theta)~[\rho_A^0 \otimes \ket{e_0}_B\bra{e_0}]^{\frac{1}{2}} ~~~\forall \xi \nn
\end{eqnarray}
where the constant $\lambda_\xi (\theta)$ can generally depend on $\xi$ and $\theta$.

We would like a condition is terms of the Kraus operators instead of the unitary U. To do this, multiply from the left by an identity in the form $I_A \otimes \sum_k \ket{e_k}_B\bra{e_k}$ to get
\begin{eqnarray}
\label{eq:cond2}
&& \sum_k [~E(\xi)^{\frac{1}{2}} \Upsilon_k'(\theta) \rho_0^{\frac{1}{2}}~]_A \otimes \ket{e_k}_B\bra{e_0} \nn \\
&& ~~~ = \lambda_\xi (\theta) ~\sum_k [~E(\xi)^{\frac{1}{2}} \Upsilon_k(\theta) \rho_0^{\frac{1}{2}}~]_A \otimes \ket{e_k}_B\bra{e_0} ~~~~ \forall \xi \nn \\
&\iff& E(\xi)^{\frac{1}{2}} \Upsilon_k'(\theta) \rho_0^{\frac{1}{2}} = \lambda_\xi (\theta) E(\xi)^{\frac{1}{2}} \Upsilon_k(\theta) \rho_0^{\frac{1}{2}} ~~\forall \xi, k
\end{eqnarray}
where the last step uses the orthogonality of $\ket{e_k}_B$. \erf{eq:cond2} defines a series of conditions that the optimal measurement must satisfy. In general, $k$ runs from $1$ to $d^2$ where $d$ is the dimension of the state-space of $\rho$, and therefore we see simply from the number of constraining equations that the optimal measurement is severely restricted.

We can reduce these two conditions to one by substituting (\ref{eq:cond2}) into the statement for the first condition (\ref{eq:cond1})
\begin{eqnarray}
\Im ~ \tr \{\sum_k \Upsilon_k\dg(\theta) E(\xi) \Upsilon_k'(\theta) ~\rho_0\} = 0 ~~~ \forall \xi \nn \\ \Rightarrow \Im ~ \tr \{\sum_k \Upsilon_k\dg(\theta) E(\xi) \lambda_\xi (\theta) \Upsilon_k(\theta) ~\rho_0\} = 0 ~~~ \forall \xi \nn \\
\Rightarrow \Im ~ \lambda_\xi (\theta) \tr \rho(\theta) E(\xi) = 0 ~~~ \forall \xi
\end{eqnarray}
Now, because the trace on the last line is always real, this condition is met if and only if $\lambda_\xi (\theta)$ is real. Therefore to summarize, the optimal measurement scheme must satisfy the conditions:
\begin{equation}
\label{eq:optmeas_cond}
E(\xi)^{\frac{1}{2}} \Upsilon_k'(\theta) \rho_0^{\frac{1}{2}} = \lambda_\xi (\theta) E(\xi)^{\frac{1}{2}} \Upsilon_k(\theta) \rho_0^{\frac{1}{2}} ~~~~\forall \xi, k
\end{equation}
where $\lambda_\xi (\theta)$ is a \textit{real} number that can depend on $\xi$.

Condition (\ref{eq:optmeas_cond}) is a condition on the optimal measurement POVM and the optimal input state. However, although it defines the optimal strategy, it is not a constructive condition. Except for special cases (that we will outline below) it is difficult to define the optimal measurement in terms of $\{\Upsilon_k\}$ and $\rho_0$ from the above condition. We will say more about the satisfiability of these conditions, and thus the achievability of the Fisher information bound below.

We conclude this subsection by noting that an immediate consequence of the form $C_\Upsilon(\theta)$ is that a pure input state, $\rho_0 \equiv \ket{\psi_0}\bra{\psi_0}$, is optimal; this follows from the linearity of trace, the concavity of density operators, and the positivity of the operator $\Pi \equiv \sum_k {\Upsilon_k\dg}'(\theta) \Upsilon_k '(\theta)$.

\subsection{Uniqueness and achievability}
\label{sec:kraus}
We derived a bound on the Fisher information above, and here we will discuss the uniqueness and achievability of this bound. These are both important questions because the achievability makes the bound meaningful (and means that $C_\Upsilon(\theta)$ can be viewed as a metric on the curve formed by the parametrized output family) and the uniqueness makes it useful as a characterization of optimality.

The Fisher information bound we have in \erf{eq:c_up} is non-unique because the Kraus operators of the quantum operation are not unique. For each choice of Kraus decomposition, the value of $C_\Upsilon(\theta)$ provides a possibly different upper bound to the Fisher information. So a natural question is: how does changing the Kraus decomposition modify the bound?

Let $\{\Upsilon_k(\theta)\}$, and $\{\Omega_k(\theta)\}$ both be Kraus operator sets for $\mathcal{E}_\theta$. Then from section \ref{sec:qops} we know that these two sets are related by a unitary transformation:
\begin{equation}
\Omega_k(\theta) = \sum_j u_{jk}(\theta)\Upsilon_j(\theta)
\end{equation}
where $u_{jk}(\theta)$ are the elements of a unitary matrix, which crucially, can also depend on $\theta$. Now, the Fisher information bound, $C_\Omega(\theta)$ for the Kraus operator choice $\{\Omega_k(\theta)\}$ in terms of the Kraus operators $\{\Upsilon_k(\theta)\}$ is:
\begin{eqnarray}
C_\Omega (\theta) &=& 4 \tr \{ \sum_k {\Omega_k\dg}'(\theta) \Omega_k '(\theta) ~\rho_0 \} \nn \\
&=& 4 \sum_k \tr[ \big( \sum_j u_{kj}^* {'}  \Upsilon_j{\dg} + \sum_j u_{kj}^* \Upsilon_j{\dg}' \big) \big(\sum_l u_{kl}' \Upsilon_l + \sum_l u_{kl}\Upsilon_l' \big) \rho_0] \nn \\
&=& 4 (~ \sum_{jkl} u_{kj}^*u_{kl} \tr[ \Upsilon_j{\dg}' \Upsilon_l' \rho_0] + \sum_{jkl} u_{kj}^* {'} u_{kl}' \tr[ \Upsilon_j{\dg} \Upsilon_l \rho_0] \nn \\
&& ~~~ + \sum_{jkl} u_{kj}^* {'} u_{kl} \tr[ \Upsilon_j{\dg} \Upsilon_l' \rho_0] + \sum_{jkl} u_{kj}^*u_{kl}' \tr[ \Upsilon_j{\dg}' \Upsilon_l \rho_0] ~)
\label{eq:fish_exp}
\end{eqnarray}
From this expression it is clear that this bound, $C_\Omega$, can be made as large as desired by appropriately choosing the unitary matrix $[u_{ij}(\theta)]$. In particular, it can be made to diverge by choosing a $[u_{ij}(\theta)]$ that is discontinuous in $\theta$. Therefore, the sensible thing to consider is the minimum value of this bound -- that is, the minimum of \erf{eq:c_up} with respect to a choice of Kraus operators. 

The second issue to be addressed is the achievability of the bound. Let the Kraus decomposition be fixed, then to show the attainability of the bound, we must show that there always exists some POVM $\{E(\xi)\}$ that can meet the optimality conditions of \erf{eq:optmeas_cond}. This attainability is a subtle task, as was pointed out by Barndorff-Nielsen and Gill in Ref. \cite{Bar.Gil-2000}. They show that in general, the optimal choice of POVM is dependent on the actual value of the unknown parameter $\theta$; which means that in practice, an adaptive strategy that narrows in on the value of $\theta$ has to be used \cite{Bar.Gil-2000}. Only in some special cases \cite{Bar.Gil.etal-2003} does one strategy achieve the bound uniformly over $\theta$. We will refer to an estimation strategy that is optimal at some value of $\theta$, but possibly not at other values of $\theta$ as a \textit{locally optimal} strategy.

Therefore to give the bound $C_\Upsilon(\theta)$ a unique meaning, we would ideally like to identify the Kraus decomposition that minimizes the bound, and show that the optimal measurement conditions containing operators from this decomposition can be met by some POVM. That is, we want to minimize $C_\Upsilon(\theta)$ over the Kraus operators for a channel, while stipulating that the optimality conditions \erf{eq:optmeas_cond} are met (locally, or globally). At this stage we do not have a method of performing this optimization for the general case, but in the next section we carry it out for a special family of quantum channels.

\subsection{A special case: the quasi-classical process}
\label{sec:sp_case}
In this section, we identify a special case where the issues of uniqueness (or equivalently, minimality) and attainability of the bound in \erf{eq:c_up} can be settled. For this case, we are able to show that the appropriate Kraus operators to use in calculating $C_\Upsilon(\theta)$ are ones forming the canonical Kraus decomposition induced by the fixed input state. 

Consider a one-parameter quantum process $\mathcal{E}_\theta$ and input state $\rho_0 = \ket{\psi_0}\bra{\psi_0}$ with canonical decomposition $\mathcal{E}_\theta(\rho_0) = \sum_k \Upsilon_k(\theta) \rho_0 \Upsilon_k\dg(\theta)$. As mentioned in \ref{sec:qops} this decomposition is characterized by the orthogonality of the Kraus operators according to an inner product based on the particular state $\rho_0$: $\mathcal{E}_\theta(\rho_0) = \sum_k \Upsilon_k(\theta) \rho_0 \Upsilon_k\dg(\theta)$ with $\tr( \Upsilon_k \dg(\theta) \Upsilon_j(\theta) \rho_0) = \delta_{jk} p_k$. Now, assume an additional orthogonality constraint on the canonical decomposition: $\tr(\Upsilon_j\dg(\theta) \Upsilon_k ' (\theta) \rho_0) = \mu_k(\theta) \delta_{jk}$, $\mu_k \in \mathbb{R}$. In terms of the local basis on the curve this constraint can be written as $\langle e_j(\theta) \vert \partial_\theta e_k(\theta) \rangle = \mu_k(\theta) \delta_{jk}$, which implies $\langle f_j(\theta) \vert \partial_\theta f_k(\theta) \rangle = 
\tilde{\mu}_k(\theta) \delta_{jk}$ where $\tilde{\mu}_k \in \mathbb{R}$ also. Note that this additional constraint specifies a \textit{quasi-classical model} where the eigenbasis of the output density operator remains the same for all $\theta$, and it is only the eigenvalues that change along the curve -- i.e. a locally orthogonal basis can also be considered a globally orthogonal basis along the curve. The output density operators $\{\rho(\theta)\}_\theta$ form a commuting parametric family. In the following, we prove the attainability and uniqueness of the Fisher information bound for this special case.

\begin{attainability}
Attainability: For the quasi-classical model, the optimality conditions
\begin{equation}
E(\xi)^{\frac{1}{2}} \Upsilon_k'(\theta) \rho_0^{\frac{1}{2}} = \lambda_\xi (\theta) E(\xi)^{\frac{1}{2}} \Upsilon_k(\theta) \rho_0^{\frac{1}{2}} ~~~~\forall \xi, k
\end{equation}
where $\{\Upsilon_k\}$ are members of the canonical decomposition can be met with a locally optimal strategy.
\end{attainability}
Proof: We will prove this by explicitly constructing the POVM that meets these conditions. Consider the optimality conditions when the input state is a pure state $\ket{\psi_0}$:
\begin{eqnarray}
E_j \Upsilon_k(\theta)'\ket{\psi_0} = \lambda_j(\theta) E_j \Upsilon_k(\theta) \ket{\psi_0} \nn \\
E_j \ket{\partial_\theta e_k} = \lambda_j(\theta) E_j \ket{e_k} \nn \\
E_j \left( \frac{p_k'(\theta)}{2\sqrt{p_k(\theta)}}\ket{f_k(\theta)} + \sqrt{p_k(\theta)}\ket{\partial_\theta f_k(\theta)} \right) \nn \\
= \lambda_j(\theta)\sqrt{p_k(\theta)} E_j \ket{f_k(\theta)}
\end{eqnarray}
for all $j, k$ (we are now assuming that the POVM has a discrete number of elements and are thus using a discrete index $j$). Now consider the choice $E_j = \ket{f_j(\theta)}\bra{f_j(\theta)}$ for the POVM -- the completeness condition on POVMs is automatically satisfied because $\{\ket{f_j}\}$ are eigenstates of a Hermitian operator and therefore span the space. Given this choice of POVM, the optimality conditions become:
\begin{eqnarray}
\left( \frac{p_k'}{2\sqrt{p_k}} + \sqrt{p_k}\tilde{\mu}_k - \lambda_j\sqrt{p_k} \right) \ket{f_k} \delta_{jk} = 0
\end{eqnarray}
where we have suppressed the $\theta$ because all quantities depend on it. This condition can be satisfied for all $j, k$ by the choice $\lambda_k = \tilde{\mu}_k + p_k '/2p_k = \mu_k/p_k$, and thus in this special case we can construct the optimal POVM. However, note that this choice, $E_j = \ket{f_j(\theta)}\bra{f_j(\theta)}$, presumes knowledge of $\theta$ and therefore can only be implemented adaptively \cite{Bar.Gil-2000}. As mentioned above, given the set $\{\ket{f_j(\tilde{\theta})}\}$ for some $\tilde{\theta}$, the additional constraint, $\langle f_j(\theta) \vert \partial_\theta f_k(\theta) \rangle = 
\tilde{\mu}_k(\theta) \delta_{jk}$, ensures that this set remains \textit{orthogonal} for all $\theta$. Despite this, the POVM has to be adapted during the estimation because the normalization of the elements of the set varies with $\theta$. $\Box$ 

\begin{uniqueness}
Uniqueness: For the quasi-classical model, the minimum of the Fisher information bound of \erf{eq:c_up} over the valid Kraus decompositions is achieved by the canonical decomposition.
\end{uniqueness}
Proof: Firstly, by the preceding theorem, we know that for the quasi-clasical model the optimality conditions can be satisfied for the canonical Kraus operators -- that is, some POVM set $\{E(\xi)\}$ can be found such that the conditions of \erf{eq:optmeas_cond} are satisfied for the canonical Kraus operators, and thus the bound is achievable. Now, let $C_\Upsilon$ denote the Fisher information bound when the canonical Kraus decomposition is used. We will show the value of the bound using any other Kraus decomposition -- i.e. $C_\Omega$ -- is larger than $C_\Upsilon$. Consider $C_\Omega$ as given in \erf{eq:fish_exp}. Using the identities $\sum_k u_{kj}^* u_{kl} = \delta_{jl}$ and $\tr( \Upsilon_k \dg(\theta) \Upsilon_j(\theta) \rho_0) = \delta_{jk} p_k$, we can rewrite this expression as:
\begin{eqnarray}
C_\Omega (\theta) &=& 4( ~ \sum_{j} \tr \Upsilon_j{\dg}' \Upsilon_j' \rho_0 + \sum_{jk} |u_{kj}'|^2 p_j \nn \\ && ~~~ + \sum_{jkl} u_{kj}^* {'} u_{kl} \tr[ \Upsilon_j{\dg} \Upsilon_l' \rho_0] + \sum_{jkl} u_{kj}^*u_{kl}' \tr[ \Upsilon_j{\dg}' \Upsilon_l \rho_0] ~ )
\end{eqnarray}
Note that the first term is simply the Fisher information bound for the canonical Kraus decomposition, $C_\Upsilon$, and the second term is always positive. Hence we see that the question of whether $C_\Omega \geq C_\Upsilon$ depends on the sign of the third and fourth terms; i.e. we are interested in the sign of:
\begin{equation}
G(\theta) \equiv \sum_{jkl} u_{kj}^* {'} u_{kl} \tr[ \Upsilon_j{\dg} \Upsilon_l' \rho_0] + \sum_{jkl} u_{kj}^*u_{kl}' \tr[ \Upsilon_j{\dg}' \Upsilon_l \rho_0]
\end{equation}
To determine this, use the optimality conditions of \erf{eq:optmeas_cond} to get rid of the derivatives within the trace. Explicitly, insert a resolution of identity of the form $\int d\xi E(\xi)$:
\begin{eqnarray}
G(\theta) &\equiv& \sum_{jkl} u_{kj}^* {'} u_{kl} \tr[ \Upsilon_j{\dg} \int d\xi E(\xi)~ \Upsilon_l' \rho_0] \nn \\ && ~~~ + \sum_{jkl} u_{kj}^*u_{kl}' \tr[ \Upsilon_j{\dg}' \int d\xi E(\xi) ~\Upsilon_l \rho_0] \nn \\
&=&  \int d\xi \lambda_\xi(\theta) \sum_{jl} \tr[ \Upsilon_j{\dg} E(\xi) \Upsilon_l \rho_0] \sum_k (u_{kj}^* {'} u_{kl} + u_{kj}^*u_{kl}') \nn \\
&=& 0
\end{eqnarray}
where the second line follows from the optimality conditions -- \erf{eq:optmeas_cond} -- and the third line follows from taking the derivative of the orthonormality condition on the elements of the unitary matrix $[u_{ij}(\theta)]$. Therefore,
\begin{eqnarray}
\label{eq:tight_bound}
C_\Omega (\theta) = 4\left(\sum_{j} \tr \Upsilon_j{\dg}' \Upsilon_j' \rho_0 + \sum_{jk} |u_{kj}'|^2 p_j \right) \geq C_\Upsilon(\theta) ~~~~~~~~\forall \theta, \nn
\end{eqnarray}
and the minimum of \erf{eq:c_up} for the quasi-classical model is achieved by the canonical Kraus decomposition. $\Box$

For this quasi-classical model, we can truly say that $C_\Upsilon (\theta)=F^*(\theta)$ (where $\{\Upsilon_k\}$ form the canonical decomposition) and hence we have a measure of statistical distance along the curve formed by the one-parameter family of output states. In fact, we can express this metric on the curve explicitly in terms of the local co-ordinate system set up by the Kraus decomposition:
\begin{eqnarray}
\label{eq:f_star_co}
\left( \frac{ds}{d\theta} \right)^2 &=& F^*(\theta) = 4 \sum_k \langle \partial_\theta e_k(\theta) \vert \partial_\theta e_k(\theta) \rangle \nn \\
&=& \sum_k \frac{(p_k(\theta)')^2}{p_k(\theta)} \nn \\ 
&& ~~~~+ 4 \sum_{k} p_k(\theta) \vert \langle f_k(\theta) \vert \partial_\theta f_k(\theta) \rangle \vert ^2 ~~
\end{eqnarray}

This result has operational significance. It means that if the input state and quantum channel are such that the output family is a mutually commuting one, then the optimal estimation scheme can be identified, and the statistical distinguishability be calculated easily.


\section{Examples}
\label{sec:examples}
In this section we will consider several examples to illustrate the results of the previous sections.

\subsection{Qubit depolarization channel}
\label{sec:depol}
The qubit depolarization channel is defined as
\begin{eqnarray}
\rho(p) = \mathcal{D}(\rho) &=& p\frac{\mathbb{I}}{2} +(1-p)\rho \nn \\
&=& (1-p)\rho \nn \\
&& + \frac{p}{3}(X\rho X + Y\rho Y + Z\rho Z)
\end{eqnarray}
where $\rho$ is a density matrix in a Hilbert space of dimension two, and $X, Y, Z$ are the Pauli matrices. This channel can be best understood by examining its action on the Bloch sphere representation of a qubit: it has the effect of uniformly shrinking the Bloch sphere towards the center. The parameter to estimate is the rate of this shrinking, parametrized by $0 \leq p \leq 1$.

From the definition of the channel, it is clear that $\{\rho(p)\}_p$ forms a commuting family. Thus we can use the canonical decomposition of the channel to analyze its estimation. 

The unitary invariance of this channel (spherical symmetry in the Bloch sphere picture) implies that all pure state inputs will perform identically when it comes to estimation performance. Therefore, choose $\ket{\psi_0} = \ket{0}$, the $+1$ eigenstate of $Z$. The canonical decomposition of the channel with respect to this initial state is:
\begin{eqnarray}
\Upsilon_1(p) &=& i\sqrt{\frac{p}{6}}~X + \sqrt{\frac{p}{6}}~Y \nn \\
\Upsilon_2(p) &=& \frac{q}{\sqrt{q+\frac{p}{3}}}~\mathbb{I} + \frac{\frac{p}{3}}{\sqrt{q+\frac{p}{3}}}~Z \nn \\
\Upsilon_3(p) &=& \sqrt{\frac{p}{6}}~X + i\sqrt{\frac{p}{6}}~Y \nn \\
\Upsilon_4(p) &=& - \sqrt{\frac{q\frac{p}{3}}{\frac{p}{3}+q}}~ \mathbb{I} + \sqrt{\frac{q\frac{p}{3}}{\frac{p}{3}+q}}~Z  \nn \\
\end{eqnarray}
where $q = 1-p$. The optimality conditions of \erf{eq:optmeas_cond} are easily seen to be satisfied by projective measurements onto the $Z$ basis -- i.e. the POVM $\{ \ket{0}\bra{0}, \ket{1}\bra{1}\}$. The statistical distance for this channel with input state $\ket{\psi_0} = \ket{0}$ is given by $F^*(p) = \frac{6}{p(9-6p)}$, which can be achieved uniformly by the estimation strategy of measuring each output in the $Z$ basis. Note that this bound diverges at $p=0$, but not at $p=1$. This is because at $p=1$ we still cannot distinguish perfectly between the action of the three Paulis with one qubit. 

\begin{figure}[h!]
\centering
\leavevmode
\includegraphics[scale=0.5]{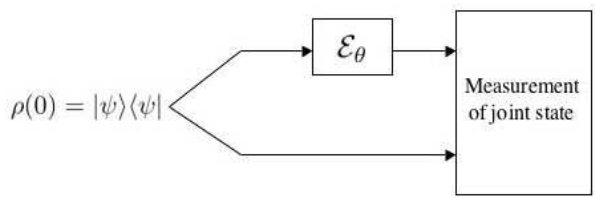}
\caption{Estimating quantum channels using entanglement. If $\mathcal{E}_\theta$ is a quantum operation acting on operators in a Hilbert space of dimension $d$, then $\dim(\ket{\psi})$ is at least $d^2$.} \label{fig:ent_est}
\end{figure}

\begin{figure}[h!]
\centering
\leavevmode
\includegraphics[scale=0.5]{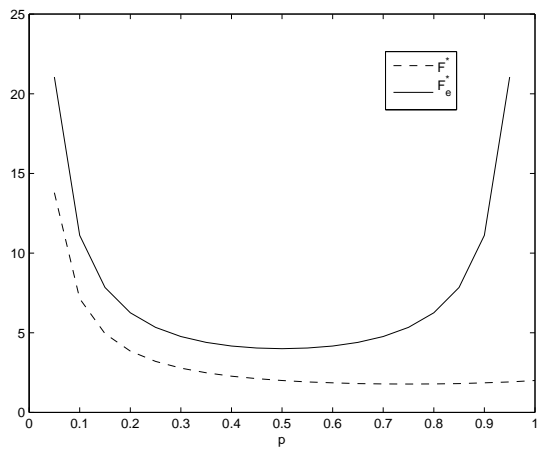}
\caption{Statistical distinguishability of the depolarizing channel with and without the use of entanglement during estimation. The values of $F^*$ at $p=0$ and $F^*_e$ at $p=0,1$ are not plotted because the quantities diverge at those points.} \label{fig:f_bounds}
\end{figure}

It is a well known fact that using entanglement can improve estimation \cite{DAr.Lo-.etal-2001, Fuj-2001}. To compare the performance of a scheme that uses entanglement to one that does not, we can compare the statistical distinguishability for the two cases. Consider the estimation of the depolarizing channel using a maximally entangled state as the input into the channel $\mathbb{I}\otimes\mathcal{D}$; see Fig. \ref{fig:ent_est}. This is a common setup for estimating channels because it can be shown that the output state completely characterizes the channel \cite{Arr.Pat-2004}. The canonical decomposition of the channel $\mathbb{I}\otimes\mathcal{D}$ with respect to the maximally entangled input state $\ket{\psi_0} = \frac{1}{\sqrt{2}}(\ket{00}+\ket{11})$ consists of the Kraus operators: 
\begin{eqnarray}
\Upsilon_1(p) &=& \sqrt{1-p}~ \mathbb{I}\otimes\mathbb{I} \nn \\
\Upsilon_2(p) &=& \sqrt{p/3}~ \mathbb{I}\otimes X \nn \\
\Upsilon_3(p) &=& \sqrt{p/3}~ \mathbb{I}\otimes Y \nn \\
\Upsilon_4(p) &=& \sqrt{p/3}~ \mathbb{I}\otimes Z.
\end{eqnarray}
Given this, it is easy to show that the optimality conditions of \erf{eq:optmeas_cond} can be satisfied by a POVM formed by projectors onto the Bell basis \footnote{The Bell states are four maximally entangled states of two qubits: $\ket{\psi^+} = \frac{1}{\sqrt{2}}(\ket{00}+\ket{11}), \ket{\psi^-} = \frac{1}{\sqrt{2}}(\ket{00}-\ket{11}), \ket{\phi^+} = \frac{1}{\sqrt{2}}(\ket{01}+\ket{10}), \ket{\phi^-} = \frac{1}{\sqrt{2}}(\ket{01}-\ket{10})$. These four states span the Hilbert space of two qubits and are therefore called the \textit{Bell basis}. The symmetric subspace of two qubit Hilbert space is spanned by the \textit{triplet} states $\ket{\psi^{\pm}}$ and $\ket{\phi^+}$, and the anti-symmetric subspace is spanned by the \textit{singlet} state $\ket{\phi^-}$.}. Note that if the singlet state, $\ket{\phi^-}$ is used as the input state, then the measurement scheme need only discriminate between the singlet and triplet subspaces to be optimal. The statistical distance in this case is $F^{*}_e(p) = \frac{1}{p(1-p)}$. Figure \ref{fig:f_bounds} plots the two values of statistical distance, and clearly shows the improved distinguishability of the parameter (uniformly across $p$) that the entanglement assisted scheme affords. 

This example illustrates the effect of using entanglement for estimation -- it can have the advantage of increasing the statistical distinguishability of channels. It also illustrates the ability of this formalism to treat entangled input states and non-local measurements. Such scenarios simply change the definition of the channel to a suitable tensor product of single party channels, while the statistical distance and optimality conditions retain their form.

\subsection{Estimation of pure $T_2$ qubit dephasing time}
\label{sec:dephasing}

There has been considerable interest recently in accurately estimating single qubit $T_2$ relaxation times for various quantum computing architectures \cite{Tyr.Lyo.etal-2003, Kha.Los.etal-2002, Vio.Aas.etal-2002}. Such estimations have also been commonplace in the NMR community for several decades now. We can use the formalism developed above to investigate the schemes used for estimating $T_2$. 

If we restrict the dynamics to be purely dephasing, we can model the single qubit channel as
\begin{equation}
\label{eq:dephasing}
\frac{d\rho}{dt} = \gamma \left( Z\rho Z - \rho \right)
\end{equation}
where $\gamma$ is the dephasing rate and the parameter we are trying to estimate. A Kraus decomposition for this process is
\begin{equation}
\rho(\theta) = \mathcal{D}_Z(\rho) = \frac{1+e^{-2\theta}}{2}~\rho + \frac{1-e^{-2\theta}}{2}~Z\rho Z
\label{eq:dephasing_kd}
\end{equation}
where $\theta = \gamma t$, is a simple transformation of the parameter we want to estimate. In the following we will notate the single qubit equal superposition states by: $\ket{+}=\frac{1}{\sqrt{2}}(\ket{0}+\ket{1})$ and $\ket{-}=\frac{1}{\sqrt{2}}(\ket{0}-\ket{1})$. Note that these are eigenstates of the operator $X$ and thus will be collectively referred to as the $X$-basis.

The standard techniques for estimating $T_2$ times are based on the \textit{spin echo} \cite{Hah-1950} pulse sequence which is essentially the preparation of a $\ket{+}$ or $\ket{-}$ initial state and then a measurement in the $X$-basis after the channel has acted. This pulse sequence actually has added features designed to nullify bulk sample inhomogeneities, but the basic idea is as mentioned. It is easy to check that given this input state, \erf{eq:dephasing_kd} is the canonical decomposition for this channel, and also that the output family $\{\rho (\theta) \}_\theta$ is a mutually commuting one. Thus we can use the operators in the canonical decomposition to determine the optimality of this scheme by checking the optimality conditions (\ref{eq:optmeas_cond}), which turn out to be
\begin{eqnarray}
(e^{-2\theta}+\lambda_j (\theta)[1+ e^{-2\theta}]) E^{1/2}_j \ket{+} = 0\nn \\
(e^{-2\theta} - \lambda_j (\theta)(1- e^{-2\theta})) E^{1/2}_j \ket{-} = 0
\end{eqnarray}
for all $j$, where we are using a discrete number of POVMs indexed by $j$. These two conditions can be met by a measurement in the $X$-basis; that is, with a POVM $\{E_+ = \ket{+}\bra{+}, E_- = \ket{-}\bra{-}\}$. The choices required for $\lambda$ are: $\lambda_+ = -1/(e^{2\theta}+1)$ and $\lambda_- = 1/(e^{2\theta}-1)$. So we see that the standard spin echo technique of estimating single qubit pure dephasing time, $T_2$, is indeed an optimal one for an X-basis input state.

Again, it is possible to show that using entanglement helps in the estimation of the parameter $\theta$ for this channel. However, a channel extension of the form $\mathbb{I}\otimes\mathcal{D}_Z$ does not increase the statistical distance, instead a channel extension of the form $\mathcal{D}_Z\otimes\mathcal{D}_Z$ must be used with a maximally entangled state.

\subsection{The random shift channel}
\label{sec:rs}
The \textit{random shift} channel is defined by the master equation
\begin{eqnarray}
\label{eq:rs_channel}
\frac{d\rho}{dt} = \gamma(U_\beta\rho U_\beta\dg - \rho) 
\end{eqnarray}
where $\gamma$ is a real, positive constant, and $U_\beta$ is a unitary operator: $U=e^{-i\beta H}$ for some continuous spectrum Hermitian operator $H$, and some real number $\beta$. A Kraus decomposition for this channel is
\begin{eqnarray}
\Lambda_k(\theta) = \frac{\theta^{k/2}}{\sqrt{k!}}~e^{-\theta/2}~ U_\beta^k ~~~~ k=0, 1, 2, ...
\label{eq:rs_channel_kd}
\end{eqnarray}
where $\theta = \gamma t$. These equations describe a channel that delivers a Poisson distributed number of unitary displacements (or unitary `kicks') by $\beta$ to the input state. The average number of kicks in a time $t$ is given by $\theta=\gamma t$, and is the parameter we are estimating. 

The way to optimally estimate the unitary version of this channel $\rho \rightarrow U_\beta\rho U_\beta\dg$ is to use the fact that $H$ is a generator of translations in some basis  \cite{Hol-2004, Bra.Cav.etal-1994}. That is, if $\ket{x}$ is an eigenstate of an operator conjugate to $H$, then $U\ket{x} = e^{-i\beta H}\ket{x} = \ket{x+\beta}$. Then the optimal scheme is to input a fiducial state $\ket{x_0}$ and to use a POVM that is formed from projectors onto translated versions of this state $\{\ket{x} ~:~ \ket{x} = U_\beta\ket{x_0}, ~\beta \in \mathbb{R}\}$ ($\langle x \vert x' \rangle = \delta(x-x')$).  For example, if $H=\hat{p}$, the momentum operator (and hence  $U$ is a spatial translation), then we would choose $E$ and $\rho_0$ to be projectors onto position eigenstates. 

Since $U_\beta$ is simply a representation of an abelian group and $U_\beta U_{\beta '} = U_{\beta+\beta '}$ we would expect the optimal scheme for estimating the random shift channel to be the same as that for estimating the unitary shift channel. Note that when the input state is a fiducial state $\ket{x_0}$, which is translated by $U_\beta$, the Kraus operators given by \erf{eq:rs_channel_kd} form the canonical decomposition. Additionally, the model is quasi-classical because the output family is a commuting one. Therefore, we can check the optimality of using the unitary channel estimation scheme for estimating the random shift channel by examining the optimality conditions with the canonical Kraus operators. $\Lambda_k(\theta)' = \left( -\frac{1}{2} + \frac{k}{2\theta} \right) \Lambda_k(\theta)$, and when the input state is $\ket{x_0}$, the optimality conditions become:
\begin{eqnarray}
\left(-\frac{1}{2}+\frac{k}{2\theta} -\lambda_\xi (\theta) \right) E^{1/2}(\xi) \Lambda_k(\theta) \ket{x_0} = 0 ~~~~\forall \xi, k 
\end{eqnarray}
Now, choosing $E(\xi) \equiv E(x) = \ket{x}\bra{x}$, we get:
\begin{eqnarray}
\left(-\frac{1}{2}+\frac{k}{2\theta} -\lambda_x \right)\delta(x-(x_0+k\beta)) \ket{x} = 0 ~~\forall x, k 
\end{eqnarray}
The left hand side is zero except when $x=x_0+k\beta$, and in that case we can choose $\lambda_x = \frac{1}{2} - \frac{x-x_0}{2\theta\beta}$ so that the left hand side goes to zero in all cases. Therefore as in the unitary case, using a POVM formed of projectors onto shifted states is optimal when the input is an element of this same set. The statistical distance for this estimation scheme is $F^*(\theta) = 1/\theta$.

\subsection{The damping channel}
\label{sec:dc}
As our final example of a one parameter quantum process, we consider the harmonic oscillator damping channel (DC). This is a quantum process described by the master equation
\begin{equation}
\label{eq:dc_master}
\frac{d\rho}{dt} = \gamma(\hat{a}\hat{\rho}\hat{a}\dg - \frac{1}{2}(\hat{a}\dg\hat{a}\rho + \rho\hat{a}\dg\hat{a} ))
\end{equation}
where $\hat{a}\dg$ and $\hat{a}$ are creation and annihilation operators for a harmonic oscillator mode, and $\gamma$ is a real, positive constant. This channel describes the effects of random photon loss. An operator sum decomposition for this process can be obtained by expanding the above master equation as a Dyson series and solving. This yields the following Kraus operators:
\begin{equation}
\label{eq:dc_kraus}
\Delta_k(\theta) = \frac{(1-e^{-\theta})^{k/2}}{\sqrt{k!}} e^{-\frac{\theta}{2} \hat{a}\dg \hat{a}} \hat{a}^k ~~~~k=0,1,2,...
\end{equation}
where $\theta = \gamma t$ is the parameter to be estimated for this channel. Note that the state space of $\rho(\theta)$ is infinite dimensional and there are also an infinite number of Kraus operators.
 
\begin{figure}[h!]
\centering
\leavevmode
\includegraphics[scale=0.6]{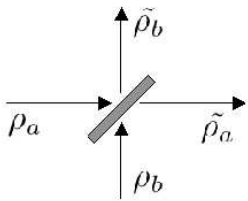}
\caption{The beam splitter interpretation of the damping channel} \label{fig:bs}
\end{figure}

One interpretation of this quantum operation is that it describes the transformation of a state when combined with the vacuum at a beam splitter (see Fig. \ref{fig:bs}). That is, the state of mode $a$ after the beam splitter is given by
\begin{equation}
\label{eq:bs}
\tilde{\rho_a} = \tr_b { U(\phi) (\rho_a\otimes\ket{0}_b\bra{0}) U\dg(\phi) } \nn
\end{equation}
where $U(\phi) = \exp( -i \phi (\hat{a}\dg\hat{b} + \hat{a}\hat{b}\dg) )$ is the beam splitter unitary transformation with $\hat{a}$ and $\hat{b}$ being the annihilation operators for modes $a$ and $b$ respectively. Evaluating this trace gives the same CP map as the damping channel with $e^{-\theta}$ replaced by $\cos ^2 \phi$, the intensity transmittance of the beam splitter. Therefore our estimation task is equivalent to the estimation of the transmittance of a beam splitter.

A common method for probing such a channel would be with a Fock (photon number) state $\ket{\psi_0} = \ket{N}$, where $N=0, 1, 2, ...$ is the number of photons in the mode. Simlarly, common measurement techniques at the channel output would be photodetection, heterodyne, or homodyne measurements. 

To determine the optimal measurement strategy when photon number states are used as input, we again confirm that firstly, the Kraus decomposition defined by \erf{eq:dc_kraus} is indeed the canonical one when number states are used, and secondly, that the channel is quasi-classical with such input states. Therefore we can decide on the optimal POVM by looking at the optimality conditions of \erf{eq:optmeas_cond}. $\Delta_k '(\theta) = \left( \frac{ke^{-\theta}}{2(1-e^{-\theta})} - \frac{1}{2}\hat{a}\dg\hat{a} \right)\Delta_k(\theta)$, and hence the conditions become:
\begin{eqnarray}
\label{eq:dc_cond}
&& \frac{ke^{-\theta}}{2(1-e^{-\theta})} E^{1/2}(\xi) \Delta_k(\theta) \ket{N} - \frac{1}{2} E^{1/2}(\xi) \hat{a}\dg\hat{a} \Delta_k(\theta) \ket{N} \nn \\
&& ~~~~~~~~~~~~ = \lambda_\xi (\theta) E^{1/2}(\xi) \Delta_k(\theta) \ket{N} 
\end{eqnarray}
for all $\xi$ and $k$. We can simplify this by applying all operators except $E(\xi)$ to the input state. Note that for $k>N$ the application of $\Delta_k(\theta)$ yields zero and the condition is trivially satisfied. For $k\leq N$ we have the condition that for all $\xi$:
\begin{eqnarray}
\left( \frac{ke^{-\theta}}{2(1-e^{-\theta})} - \frac{N-k}{2} - \lambda_\xi (\theta) \right) E^{1/2}(\xi) \ket{N-k} = 0  
\end{eqnarray}
To satisfy this we can choose $E(\xi) \equiv E(M) = \ket{M}\bra{M}$, a number state projector, which corresponds to photodetection. This choice requires $\lambda_M = \frac{Ne^{-\theta}-M}{2(1-e^{-\theta})}$. Therefore the best strategy for estimating the damping channel when Fock states are used as input is to perform photodetection at the output. The statistical distance for this scenario is $F^*_N (\theta) = \frac{N}{e^{\theta}-1}$, and hence larger $N$ at the input makes the process more distinguishable. 

\subsection{A comment on estimation}
\label{sec:est_comm}
In all the examples considered above, except the random shift channel, we are trying to estimate a continuous parameter with experiments that have discrete outcomes. This may seem peculiar, but the situation is clarified by the observation that the parameter is always a continuous function of the probabilities of the discrete outcomes (or rather, the probabilities are functions of the parameter). Therefore from the point of view of the estimator,  the problem is the same as estimating the parameter of a probability distribution from independent experiments that sample that distribution. This is a well known estimation problem in classical estimation theory and a \textit{maximum likelihood estimator} \cite{Nah-1969} would be a practical estimator that would also achieve the Fisher information bound asymptotically. 

\section{Conclusion}
\label{sec:concl}

We have investigated the problem of optimally estimating a general one parameter quantum process. We have attempted to obtain characterizations of estimation optimality in terms of a common representation of quantum processes, the Kraus decomposition. We derived a bound on estimation accuracy and conditions of optimality when the input state is fixed, however, the non-uniqueness of the Kraus decomposition causes this bound to be non-unique. It also makes proving the achievability of the bound difficult. However, we have shown that in the special case of a quasi-classical channel, the issues of uniqueness and attainability can be settled, and in this special case, the characterization of optimal estimation (with a fixed input state) we derived is useful in determining the statistical distinguishability of quantum processes. 

Representing a quantum process in terms of its Kraus decomposition has the advantage that it is often easy to do, however, in view of the above treatment, this representation is difficult for characterizing optimal estimation strategies because of its non-uniqueness. The immediate direction in which this work could be extended is to investigate the possibility of settling the issues of attainability and uniqueness for a general quantum channel. In particular, explicit expressions for the optimal POVM from the optimality conditions \erf{eq:optmeas_cond}, for a fixed set of Kraus operators, would be extremely useful.

\section{Acknowledgements}
\label{sec:acks} 
We gratefully acknowledge the support of the Australian Research Council Centre of Excellence in Quantum Computer Technology. We thank Lajos Diosi and the referees for pointing out errors in the first draft and several useful comments. MS would also like to acknowledge useful discussions with Duncan Mortimer, Michael Nielsen, Charles Hill, and Carl Caves.
 
\section*{References}
\bibliography{mybib}

\begin{thebibliography}{10}

\bibitem{Hel-1976}
C.~V. Helstrom.
\newblock {\em Quantum Detection and Estimation Theory}.
\newblock Academic Press, New York, 1976.

\bibitem{Hol-1982}
A.~S. Holevo.
\newblock {\em Probabilistic and statistical aspects of Quantum Theory}.
\newblock North-Holland, Amsterdam, 1982.

\bibitem{Bra.Cav-1994}
S.~Braunstein and C.~Caves.
\newblock Statistical distance and the geometry of quantum states.
\newblock {\em Phys. Rev. Lett.}, 72:3439, 1994.

\bibitem{Fuj.Nag-1995}
A.~Fujiwara and H.~Nagaoka.
\newblock Quantum fisher metric and estimation for pure state models.
\newblock {\em Phys. Lett. A}, 201:119, 1995.

\bibitem{Bar.Gil-2000}
O.~E. Barndorff-Nielsen and R.~D. Gill.
\newblock Fisher information in quantum statistics.
\newblock {\em J. Phys. A}, 30:4481, 2000.

\bibitem{Gil.Mas-2000}
Richard~D. Gill and Serge Massar.
\newblock State estimation for large ensembles.
\newblock {\em Phys. Rev. A}, 61:042312, 2000.

\bibitem{Hol-2004}
A.~S. Holevo.
\newblock Asymptotic estimation of shift parameter of a quantum state.
\newblock {\em Prob. theory and appl.}, 49:207, 2004.

\bibitem{Par.Reh-2004}
Matteo Paris and Jaroslav Rehacek~(Eds.).
\newblock {\em Quantum state estimation}.
\newblock Number 649 in Lecture Notes in Physics. Springer, 2004.

\bibitem{Woo-1981}
William~K. Wootters.
\newblock Statistical distance and hilbert space.
\newblock {\em Phys. Rev. D}, 23:357, 1981.

\bibitem{Bar.Gil.etal-2003}
O.~E. Barndorff-Nielsen, R.~D. Gill, and P.~E. Jupp.
\newblock On quantum statistical inference.
\newblock {\em J. Roy. Stat. Soc. B}, 65:775, 2003.

\bibitem{Bra.Cav.etal-1994}
S.~Braunstein, C.~Caves, and G.~J. Milburn.
\newblock Generalized uncertainty relations: Theory, examples, and lorentz
  invariance.
\newblock {\em Annals of Physics}, 247:135, 1994.

\bibitem{Bal-2004}
Manuel~A. Ballester.
\newblock Estimation of unitary quantum operations.
\newblock {\em Phys. Rev. A}, 69:022303, 2004.

\bibitem{Fuj-2001}
A.~Fujiwara.
\newblock Quantum channel identification problem.
\newblock {\em Phys. Rev. A}, 63:042304, 2001.

\bibitem{Fuj.Hir-2003}
Akio Fujiwara and Imai Hiroshi.
\newblock Quantum parameter estimation of a generalized pauli channel.
\newblock {\em J. Phys. A}, 36:8093, 2003.

\bibitem{Sas.Ban.etal-2002}
Masahide Sasaki, Masashi Ban, and Stephen~M. Barnett.
\newblock Optimal parameter estimation of depolarizing channel.
\newblock {\em Phys. Rev. A}, 66:022308, 2002.

\bibitem{Cho-1975}
M.~D. Choi.
\newblock Completely positive linear maps on complex matricies.
\newblock {\em Lin. Alg. and App.}, 10:285, 1975.

\bibitem{Kra-1983}
K.~Kraus.
\newblock {\em States, Effects, and Operations: Fundamental Notions of Quantum
  Theory}, volume 190 of {\em Lecture Notes in Physics}.
\newblock Springer-Verlag, Berlin, 1983.

\bibitem{mikeandike}
M.~A. Nielsen and I.~L. Chuang.
\newblock {\em Quantum computation and quantum information}.
\newblock Cambridge University Press, 2001.

\bibitem{Nie.Cav.etal-1998}
M.~A. Nielsen, Carlton~M. Caves, Benjamin Schumacher, and Howard Barnum.
\newblock Information-theoretic approach to quantum error correction and
  reversible measurement.
\newblock In {\em Proc. Roy. Soc. London Series A}, volume 454, pages 257--486,
  1998.

\bibitem{Ben.Flo-2005}
Fabio Benatti and Roberto Floreanini.
\newblock Open quantum dynamics: Complete positivity and entanglement.
\newblock {\em Int. J. Mod. Phys.}, B19:3063, 2005.

\bibitem{Cov.Tho-1991}
T.~M. Cover and J.~A. Thomas.
\newblock {\em Elements of Information Theory}.
\newblock Wiley-Interscience, 1991.

\bibitem{Nah-1969}
Nasser~E. Nahi.
\newblock {\em Estimation theory and applications}.
\newblock John Wiley and Sons, 1969.

\bibitem{DAr.Lo-.etal-2001}
G.~Mauro D'Ariano, Paoloplacido Lo~Presti, and Matteo G.~A. Paris.
\newblock Using entanglement improves the precision of quantum measurements.
\newblock {\em Phys. Rev. Lett.}, 87:270404, 2001.

\bibitem{Arr.Pat-2004}
Pablo Arrighi and Christophe Patricot.
\newblock On quantum operations as quantum states.
\newblock {\em Annals Phys.}, 311:26, 2004.

\bibitem{Tyr.Lyo.etal-2003}
A.~M. Tyryshkin, S.~A. Lyon, A.~V. Astashkin, and A.~M. Raitsimring.
\newblock Electron spin relaxation times of phosphorus donors in silico.
\newblock {\em Phys. Rev. B}, 68:193207, 2003.

\bibitem{Kha.Los.etal-2002}
A.~V. Khaetskii, D.~Loss, and L.~Glazman.
\newblock Electron spin decoherence in quantum dots due to interaction with
  nuclei.
\newblock {\em Phys. Rev. Lett.}, 88:186802, 2002.

\bibitem{Vio.Aas.etal-2002}
D.~Vion, A.~Aassime, A.~Cottet, P.~Joyez, H.~Pothier, C.~Urbina, D.~Esteve, and
  M.~H. Devoret.
\newblock Manipulating the quantum state of an electrical circuit.
\newblock {\em Science}, 296:886, 2002.

\bibitem{Hah-1950}
E.~Hahn.
\newblock Spin echoes.
\newblock {\em Phys. Rev.}, 80:580, 1950.

\end{thebibliography}
\bibliographystyle{unsrt}

\end{document}